\documentclass{sig-alternate}

\usepackage{hyperref}
\usepackage{graphicx}

\begin{document}

\toappear{Preprint submitted to ExtremeCom 2013}

\title{Spreading huge free software without internet connection\\ via self-replicating USB keys}


\numberofauthors{1}
\author{
    \alignauthor Thierry Monteil\\
    \affaddr{LIRMM, UMR 5506, CNRS, Universit\'e Montpellier II}\\
    \affaddr{161 rue Ada, 34392 Montpellier Cedex 5, France}\\
    \email{thierry.monteil@lirmm.fr}
}

\maketitle
\begin{abstract}

We describe and discuss an affordable way to spread huge software without
relying on internet connection, via the use of self-replicating live USB keys.

\end{abstract}






\section{Introduction}

Despite claims and hopes supporting this idea, being free software is
definitely not a sufficient condition for being usable in ``developing
countries''. A reason is that free software and internet are intimately
interdependent. Internet growth was made possible through open standards and
implementations. Being free as a beer and built by communities interacting all
over the world, free software relies on the internet to be spread and
developed.

Unfortunately, the cost of bandwidth is very expensive in those countries
especially relative to the local income \cite{price}, and the available
bandwidth does not allow huge downloads making downloading a Linux distribution
(and its updates) more expensive than getting cracked copies of Windows
operating system. 
This network consideration
seems to be an important bottleneck in spreading free software where bandwidth
is a rare commodity \cite{ghabuntu}.


\begin{figure}[h]
\begin{center}
\includegraphics[scale=0.13]{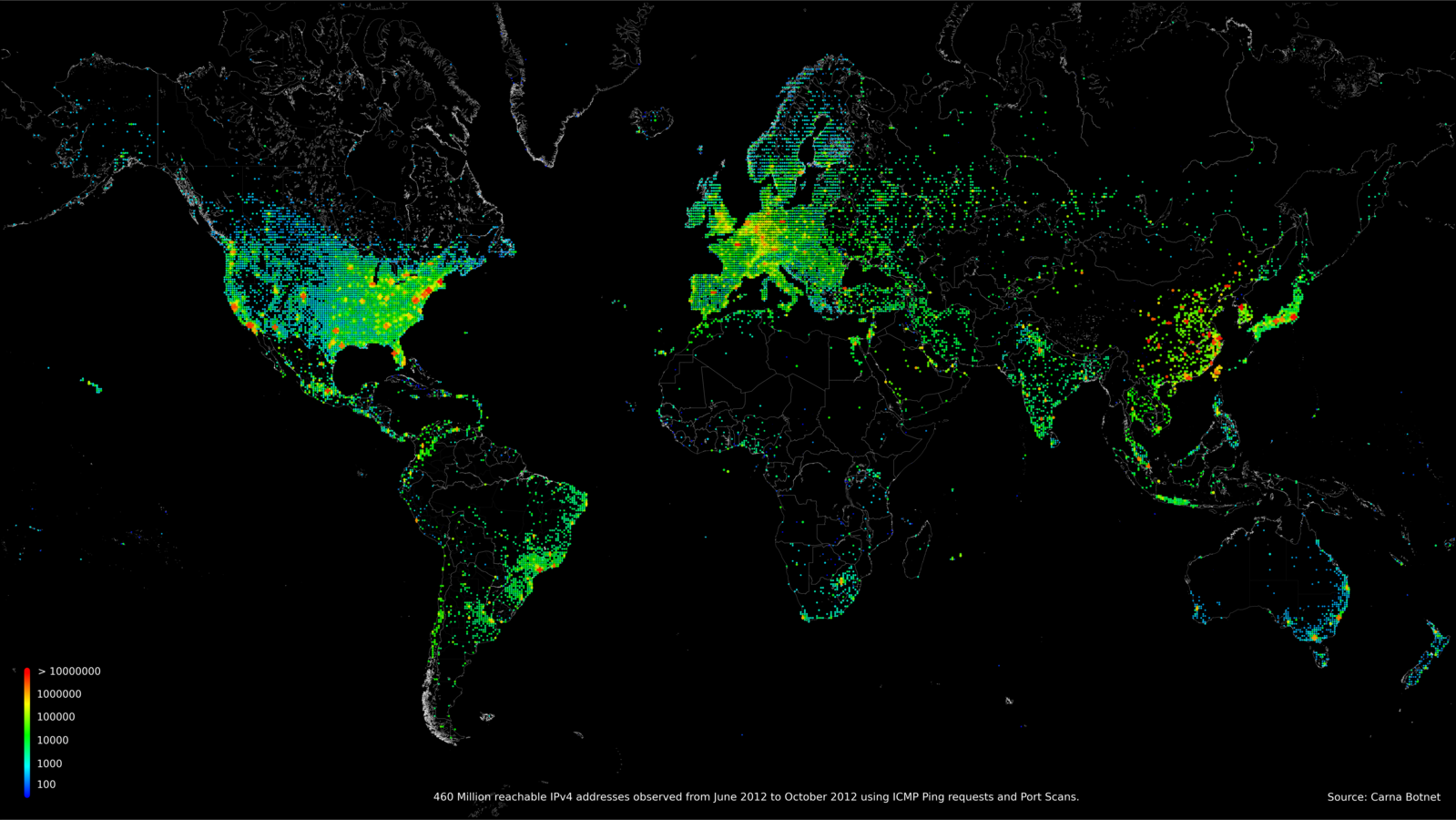}
\caption{\label{ipv4_map}World map of IPv4 addresses, from \cite{census}.}
\end{center}
\end{figure}

\begin{figure}[h]
\begin{center}
\includegraphics[scale=0.28]{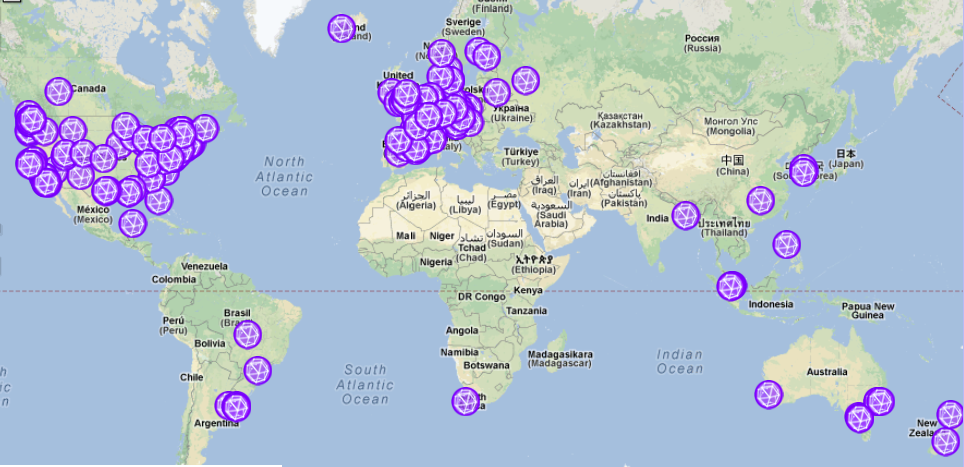}
\caption{\label{dev_map}Sage developers map, from \cite{sage}.}
\end{center}
\end{figure}
To thwart that phenomenon, Canonical used to distribute free CD of the Ubuntu
distribution, but unfortunately ended this service \cite{shipit}, forgetting in
the meantime that downloading has a cost:
\begin{quote} After delivering millions of Ubuntu CD's to millions of new users,
    our ShipIt program has finally run its course. While we can no longer
    deliver free CD's through the program, it's still easy to get Ubuntu. You
    can download Ubuntu for free from Ubuntu.com or you can buy a CD straight
from the Canonical shop.  \end{quote}
Even for a well-funded distribution, burning and sending free CD at each (6
months) release has a cost!


We try to propose here an alternative way to distribute huge free software at
low cost, among a defined community.



\section{Use case: distributing Sage mathematical software in West Africa}

\subsection{Context}

Sage \cite{sage} is a mathematical software aiming at being a free alternative
to Mathematica, Maple, Matlab, Magma,... Its compressed minimal binary weighs
about 650MB. Uncompressed binary with additional GAP database weighs up to 3GB.

In order to prepare the CIMPA/ICPAM research school to be held in 2012 that
contained a lecture about Sage \cite{cimpa}, a workshop was organized in 2011
with local mathematician colleagues \cite{days}.  This was the occasion to
distribute this software among participants and start playing with it, in order
to let the local team to help during the Sage lectures (73 participants
attended the school, which requires support in helping participants to debug
their code).

Concerning installations, we experienced quite a lot of problems related to the
lack of internet connection at the university itself, and to the age of some
machines (starting from Pentium 3). 
First, Sage does not run natively on Windows: the classical workaround is to
run its server from a virtualized Linux distribution. 
Such a method is not a viable solution on slow computers, so the software needs
to be run directly from a non-emulated Linux/BSD/Unix.
We brought copies of Linux distributions, but running Sage binaries on them
requires some small additional dependencies, a small detail that become a harsh
when they have to be downloaded from a cybercaf\'e a few kilometers from the
campus.
Moreover, repartitioning the users hard disks without backup may cause personal
data loss.
Therefore, the best solution was to run Sage from the live CD (turned into a
live USB with persistence to be able to store personal worksheets from one
session to the next one), which is not emulated, contains required
dependencies, and does not touch the user hard disk. 

Another advantage of the USB key over fast internet connection is its worldwide
availability, at low cost.

\begin{figure}[h]
\begin{center}
\includegraphics[scale=0.20]{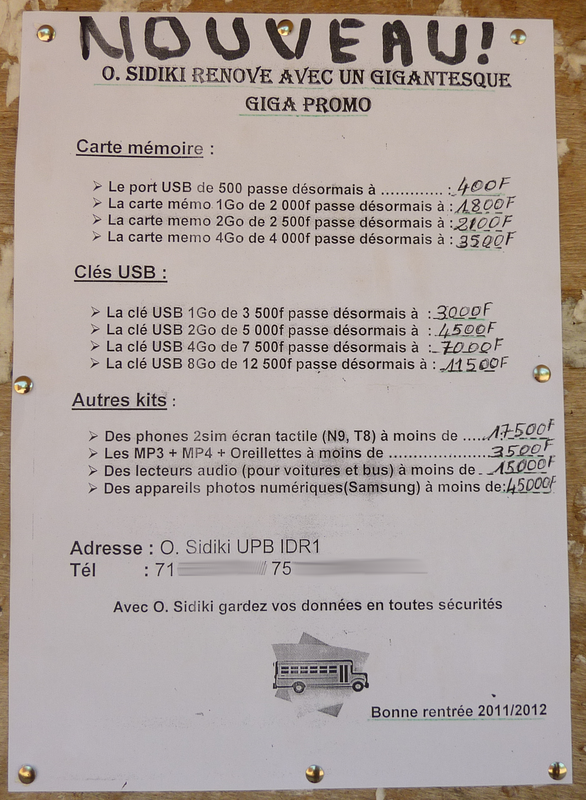}
\caption{\label{promo}Discount on USB keys.}
\end{center}
\end{figure}

The main concern with the live USB is the difficulty to spread it among
colleagues and students after the workshop as it requires from the user some
knowledge on how to make the target USB key bootable.

\subsection{Solution}

During this first workshop in 2011, we built a prototype of a script allowing
the existing Sage Live (based on Puppy Linux \cite{puppy}) to clone itself on
another USB key, indefinitely, in one click.
As a fork of an existing live USB, it turned out to be hard to maintain, hence
we decided to start an autonomous USB key based on Live Debian system
\cite{debianlive}: Sage Debian Live \url{http://sagedebianlive.metelu.net/}
\cite{sagedebianlive}.



\section{Description of the self-replicating live USB key}

The main feature of this USB key, besides running Sage under Linux, is its
ability to clone itself on any sufficiently large USB key, in a few clicks,
without requiring any knowledge.

\begin{figure}[h]
\begin{center}
    \includegraphics[scale=0.5]{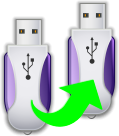}
\caption{\label{clone}``Clone the USB key'' icon}
\end{center}
\end{figure}


So, not only the Sage software is transmitted, but also the Linux distribution,
which holds the self-replicating capability.

\subsection{Features and design principles}

\subsubsection{Upgrading an existing USB key}

    When the USB key already contains some data, or a previous version of the
    software, the clone script can keep the personal data and installs the
    software over the previous version. This feature is interesting in a
    teaching environment, ensuring a uniform versioning among students, that
    ease code debugging.

\subsubsection{Offline autonomy}

    Most of the live USB distributions aim at being small (they usually try to
    fit on a CD-ROM, and sometimes much less). This diet is done by removing
    most of the documentation (for example, in Puppy Linux, the {\tt man}
    command opens a web browser to an on-line manual page, which is clever, but
    unsuitable for an off-line use), and having only a few packages installed by
    default.  Hence, being light implies being connected. Here, we need to take
    the opposite way, and try to be as exhaustive and autonomous as possible.
    Being fat is not a problem (the clone operation is a bit longer but the
    key-to-key copy bandwidth is quite high). For example, we included
    translations of some software and all locales are generated during the
    build. The current Sage Debian Live weighs 2.7 GB, which corresponds to 8.4
    GB of compressed data. It can be cloned in about 10 minutes, depending on
    the speed of the USB key.

\subsubsection{Straightforward persistence}

    The installation is made easy by the fact that the personal data are stored
    on the main {\tt vfat} partition, bypassing the traditional persistence
    scheme which requires another partition in a POSIX compliant
    format.  Since {\tt vfat} is not POSIX compliant, our straightforward
    persistence scheme bind-remounts the main partition with an additional
    POSIX layer (thanks to fuse-posixovl).  It allows the user to have a direct
    access to the data she produced with Sage (or \LaTeX ~or any other tool
    distributed with the USB key), when using the USB key in a standard way.

\subsubsection{Keeping personal data}

    Though straightforward persistence allows the user to create and store
    personal content, no personal data is duplicated to the cloned USB key. For
    this purpose, the clone script uses a white list containing the files that
    have to be copied to the target USB key. This can be made easy by the fact
    that the whole system is enclosed in a single squashfs read-only compressed
    filesystem.

\subsubsection{Sharing interesting data between users}
    
    An exception is made for data put by the user in the {\tt /share}
    directory, which will be transmitted to the next USB keys. This allows
    interaction between participants (e.g. sharing pictures), as well as the
    possibility for the source to distribute additional files (lecture notes,
    exercises sheets), without having to recompile a whole USB key image.

\subsubsection{CD bootloader}

    Old computers are sometimes not able to boot from the USB key, but all of
    them are able to boot from the CD-ROM, hence the live USB contains a small
    bootloader ISO \cite{plopkexec} which can be burnt on a CD in one click,
    allowing the USB key to be loaded after booting on the CD.

    \begin{figure}[h]
    \begin{center}
        \includegraphics[scale=0.5]{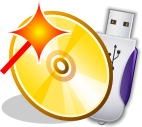}
    \caption{\label{bootloader}``Create a bootloader CD-ROM'' icon}
    \end{center}
    \end{figure}

    A not planned side effect of the ISO bootloader is the possibility to boot
    on MACs, whose BIOS is incompatible with the classical Master Boot Record
    partition table.

\subsubsection{Sublinear spread complexity}

    As noticed, the clone procedure takes about 10 minutes for a key containing
    2.7GB of data (about 8 minutes for an upgrade), which corresponds to a
    bandwidth of 4.5MB/s (this measurement was done with the cheapest 8GB USB
    2.0 key we bought).
    Since each cloned USB key can become a new seeder, the live USB key can be
    spread among $n$ participants in time $O(log(n))$, which is not possible
    via classical wireless hubs, even with a locally-hosted mirror.
    In our case, since at least 3 different versions (bug fixes, improvements
    and user feedback) were deployed during the 2 weeks school in Bobo
    Dioulasso, this \emph{sublinear spread complexity} was of great interest,
    since the key could be redeployed among participants in two coffee breaks.
    In a room of 60 participants, there will be 6 cloning rounds until
    everybody gets a fulfilled key, so even if participants take 5 minutes to
    boot and play around before launching the clone, the amortized bandwidth
    becomes 30MB/s.
    This bandwidth can still be increased by a constant factor, by allowing a
    USB key to spawn as many USB keys as the number of USB ports available.
    However, we have to be very careful in the targets selection since cloning
    the USB key on the hard disk accidentally would be fatal (the current tests
    are quite strict, and refuse to start if the total number of USB devices is
    not equal to 2, hence if for some reason, one of the internal devices is
    recognized as a USB drive, the script will not go further. This happened on
    one laptop, though the extra USB device was the internal CD-ROM reader).

\subsubsection{Modularity}

    The source code is made of modules, aiming at being as independent as
    possible. In particular, all the Sage stuff can be easily removed, to
    spread any kind of software, with respect to the limitations described in
    \ref{integrity}.

\subsection{Limitations}

\subsubsection{Getting feedback}

    It is interesting to see how could such a USB key be spread and, in case it
    is used to share files within a community, what is the data renewal or the
    mixing rate.  Which feature offers an ``evolutive advantage'' to the USB
    key?  For example, can the USB key cross borders between countries not
    sharing the same language ?
    Software statistics are usually done via counting the number of downloads,
    but here we precisely want to measure the off-line spreading of the USB
    key, not the initiation of a stream.
    For this, we built a genealogy system that allows to track the history of
    the ancestors of the USB key as well as which USB keys it spawned.

\begin{figure}[h]
\begin{center}
\tiny
\begin{verbatim}
    p Sage 5.6 Debian Live beta4 2013-01-25 en_US.UTF-8 - wheezy - 686-pae
    i 2013-02-17 13:01:38+00:00 - fr_FR.UTF-8 - 4023296 -          1
    s 2013-02-17 13:15:19+00:00 - fr_FR.UTF-8 - 4023296 -          1
    s 2013-02-17 13:31:36+00:00 - fr_FR.UTF-8 - 4023296 -          1
    s 2013-02-17 14:40:22+00:00 - fr_FR.UTF-8 - 4023296 -          1
    s 2013-02-19 15:56:46+00:00 - fr_FR.UTF-8 - 4023296 -          1
    s 2013-02-19 16:19:28+00:00 - fr_FR.UTF-8 - 4023296 -          1
    s 2013-02-19 16:39:30+00:00 - fr_FR.UTF-8 - 4023296 -          1
    s 2013-02-19 16:47:17+00:00 - fr_FR.UTF-8 - 4023296 -          1
    u p Sage 5.8 Debian Live ejcim 2013-04-06 fr_FR.UTF-8 - wheezy - 686-pae
        i 2013-04-07 08:44:13+00:00 - fr_FR.UTF-8 - 7692288 -          1
        s 2013-04-07 08:51:52+00:00 - fr_FR.UTF-8 - 7692288 -          1
        s 2013-04-07 11:58:04+00:00 - fr_FR.UTF-8 - 7692288 -          1
      i 2013-04-07 11:58:04+00:00 - fr_FR.UTF-8 - 7692288 -          1
      s 2013-04-08 07:48:27+00:00 - fr_FR.UTF-8 - 7692288 -          1
      u Sage 5.8 Debian Live beta4 2013-04-26 en_US.UTF-8 - wheezy - 686-pae
        i 2013-04-26 16:09:12+00:00 - en_US.UTF-8 - 7692288 -          1
      u p Sage 5.9 Debian wheezy Live 3.0.5-1 2013-05-09 en_US.UTF-8 - wheezy - 686-pae
          i 2013-06-16 19:17:25+00:00 - en_US.UTF-8 - 7692288 -          1
          s 2013-06-16 19:26:52+00:00 - en_US.UTF-8 - 7692288 -          1
        i 2013-06-16 19:26:52+00:00 - en_US.UTF-8 - 7692288 -          1
      s 2013-06-17 07:15:42+00:00 - en_US.UTF-8 - 7692288 -          1
\end{verbatim}
\caption{\label{genealogy}Genealogy of a key}
\end{center}
\end{figure}

    To avoid privacy leaks (and annoying pop-up when the user get connected to
    the internet), the genealogy file is sent only when the user explicitly
    decides to, and actually, the associated contact form was only used as a
    way to get direct support (a non-planned feature).

    \begin{figure}[h]
    \begin{center}
        \includegraphics[scale=0.20]{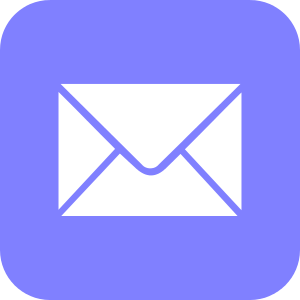}
    \caption{\label{contact}``Contact the author'' icon}
    \end{center}
    \end{figure}

%

\subsubsection{Integrity}\label{integrity}

The self-replicating scheme can be compared with a software virus. And indeed,
the live USB key can be easily modified by an attacker, by simply modifying the
publicly available sources and bootstrap her custom USB key.
The issue is that the USB key cannot test its own integrity: an attacker could
distribute a USB key with modified versions of all tools allowing integrity
checks (signing or checksumming software).
Hence, to get confident that the USB key was not modified during the
transportation, the user has to check the SHA256SUMS on the website, with her
own tools. This requires an internet connection, some knowledge and some
motivation.

Therefore, this communication-via-replication protocol currently relies on
trust, and should be only used for short distance communication (e.g. in a
classroom deployment, there is not enough time to modify the USB key between
two consecutive clone operations), or among a small structured community.
This is well adapted for a specialized software such as Sage, probably less
for a mainstream software such as Linux.

For example, one could imagine spreading off-line versions of Wikipedia, using
Kiwix software, but in the current framework, this could have harmful
consequences.

\subsubsection{Ecological impact}

The size of the USB keys is still growing very fast, making them rapidly
obsolete. In relying upon them, one promotes their consumption and disposal.

\section{Related approaches}

\subsection{Key System hardware}

This company sells USB keys with both male and female USB connectors, allowing
the data to be directly transferred to another USB key, without the need for a
computer \cite{key2key}. We do not know whether is is able to format the target
USB key though.

While being an appealing solution, this requires a specific hardware, which
should be spread along with the software if we want the diffusion to be
iterated. It is moreover much more expensive than standard USB keys (50 euros
for a 4GB USB key as of 2013).

\subsection{USB nets}

In the previous Extremecom conference, Panayotis Antoniadis, Larch Chen and
Franck Legendre, studied USB nets (see \cite{usbnet} and references therein),
where participants distribute USB keys filled with interesting data.

The main issue addressed in the paper was in motivating the community to behave
in an altruist way. In aiming to change our cultural mind via the use of some
technology, this is quite an ambitious project.

Here, the hardware is not transferred, only the data is. Each participant to
the network needs to have a single USB key, and the motivation for a user to
buy a USB key is actually selfish. In particular, privacy leaks related to the
recovery of deleted files is avoided since the USB keys are not assumed to
change their owner.

That said, one could imagine a USB net based on self-replicating live USB keys:
the interesting data could be put on the {\tt /share} directory, and the live
system could be limited to various media readers, as well as a tool for
indexing and organizing the data (e.g via an off-line wiki or a collaborative
spreadsheet). A restriction for this being to limit the spread to small
communities, as explained in Subsection \ref{integrity}.

Conversely, the fun introduced in USB nets to get feedback could be used in our
case.

\subsection{Comparison with other live USB}

Technically, most live USB have the possibility to replicate themselves,
provided they contain partitioning tools (like {\tt sfdisk} or {\tt parted}),
formatting tools (like {\tt mkfs.vfat}), and bootloaders (like {\tt syslinux}),
which are all small and standards binaries. Writing a clone script is not hard
either.


It seems that none of the existing live USB aims at being spread via a
key-to-key protocol, they are mainly built from a CD ISO file, using either a
software like unetbootin or yumi or using the isohybrid format, allowing the
CD ISO image to be directly copied to the USB key. Here, the live USB is the
final target.

%
%
%

The closest tools we found in inspecting existing solutions were the two
install wizards distributed with Puppy Linux. Unfortunately, they are somehow
too powerful and can not be used out of the box for an inexperienced Linux
user. If we start the process with an unformatted USB key, the wizard will
suggest to use the ``Universal Installer'', then propose to select the target
device (among which the source USB key appears!), then propose the user to use
{\tt gpared} to configure its partitions, this operation requires the user to
take an initiative. The second tool named ``Bootflash USB installer'', will
format and make the USB key bootable in one click, but will require the user to
provide an ISO file as a source. At this point, it is possible to relaunch the
first wizard and finish the installation, but most Linux newcomer will
surrender before that step.

One reason for such a complex behaviour may be the fact that the tool is aimed
at installing the distribution, no matter the target disk (they all belong to a
single list).
There is however quite a big difference between replicating the USB key to
another one and installing its content to a computer hard disk.
Physically, the first copies a squashfs image to the target USB key, the second
should do an uncompressed install with taking care about partitions, other
present OSes, swap, etc.
Semantically, the action of replicating the USB key corresponds to circulating
the software with as few modifications as possible, whereas hard disk
installation corresponds to settle it in a cleaner way.
Those two processes should be separated.

By the way, it is worth noticing that, though the possibility to install the
live USB key on the computer hard disk is available in the source code of the
Sage Debian Live key, is currently not enabled in the distributed keys to avoid
misuses and possible data loss.

\section{Conclusion}

The introduction of a self-replicating bootable USB key allowed the Sage
lectures of the CIMPA/ ICPAM research school to go smoothly, as well as some
further Sage deployments.

While being very efficient for our purpose of spreading huge free software
inside a community, it is definitely not advisable for spreading sensitive
software, or for large-scale distribution.

A possible workaround could be to take advantage of the upgrade feature, which
creates some recurrence in the interactions between USB keys, possibly allowing
some trust accumulation. For example, if a user has two live USB keys, she can
expose the first one to be upgraded by someone else, and then boot on the
second live USB key to check the integrity of the first one. In that checking
process, we should ensure that the new files are authenticated, but also that
the boot sector really points to those files (not to a malicious system located
somewhere else on the filesystem).
When users only have one live USB key, we could imagine a game where, when two
users meet each other, instead of booting on the newest USB key to upgrade the
oldest USB key, the users could randomly draw which USB key should boot in
order to check the other one for integrity.

Such protocols still needs to be evaluated and experimented, they should be
understandable by anyone and should not provide a false sense of security when
misused.


\section{Acknowledgments}

I would like to thank 

\begin{itemize}

    \item all participants of both the preliminary workshop and the CIMPA/ICPAM
        research school, for their feedback and encouragements in the first
        versions of the USB key. 

\item the GDR-IM of the CNRS, for financing the purchase of USB keys for the
    participants of the research school.

\item Emil Widmann, who built the Puppy-based Sage Live CD, allowing me to
    start from somewhere.

\item Special dedication to Sokous Som\'e.

\end{itemize}

\bibliographystyle{abbrv}

\bibliography{liveusb}  

\end{document}